\documentclass[twocolumn,amsmath,prl,nobibnotes,showpacs,superscriptaddress]{revtex4}

\usepackage{graphicx,mathrsfs,amssymb}
\usepackage{dcolumn}
\usepackage{bm}
\usepackage{longtable}

\newcommand{\bra}[1]{\ensuremath{\left< #1 \right|}}
\newcommand{\ket}[1]{\ensuremath{\left| #1 \right>}}

\begin{document}
\title{Experimental nonclassicality of single-photon-added thermal light states}

\author{Alessandro Zavatta}
\email{azavatta@inoa.it}\affiliation{Istituto Nazionale di Ottica Applicata (CNR), L.go E. Fermi, 6,
I-50125, Florence, Italy} \affiliation{Department of Physics, University of Florence, I-50019 Sesto
Fiorentino, Florence, Italy}

\author{Valentina Parigi}
\affiliation{Department of Physics, University of Florence, I-50019 Sesto Fiorentino, Florence,
Italy} \affiliation{LENS, Via Nello Carrara 1, 50019 Sesto Fiorentino, Florence, Italy}

\author{Marco Bellini}
\email{bellini@inoa.it} \affiliation{Istituto Nazionale di Ottica Applicata (CNR), L.go E. Fermi,
6, I-50125, Florence, Italy} \affiliation{LENS, Via Nello Carrara 1, 50019 Sesto Fiorentino,
Florence, Italy}

\date{\today}
\begin{abstract}
We report the experimental realization and tomographic analysis of novel quantum light states obtained
by exciting a classical thermal field by a single photon. Such states, although completely incoherent,
possess a tunable degree of quantumness which is here exploited to put to a stringent experimental test
some of the criteria proposed for the proof and the measurement of state nonclassicality. The quantum
character of the states is also given in quantum information terms by evaluating the amount of
entanglement that they can produce.
\end{abstract}


\maketitle

\section{Introduction}
The definition and the measurement of the nonclassicality of a quantum light state is a hot and
widely discussed topic in the physics community; nonclassical light is the starting point for
generating even more nonclassical states~\cite{lund04,jeong05} or producing the entanglement which
is essential to implement quantum information protocols with continuous
variables~\cite{kim02,braunstein05}. A quantum state is said to be nonclassical when it cannot be
written as a mixture of coherent states. In terms of the Glauber-Sudarshan $P$
representation~\cite{glauber63:131,sudarshan63}, the $P$ function of a nonclassical state is highly
singular or not positive, i.e. it cannot be interpreted as a classical probability distribution. In
general however, since the $P$ function can be badly behaved, it cannot be connected to any
observable quantity. In recent years, a nonclassicality criterion based on the measurable
quadrature distributions obtained from homodyne detection has been proposed by Richter and
Vogel~\cite{vogel00}. Moreover, a variety of nonclassical states has recently been characterized by
means of the negativeness of their Wigner function~\cite{lvovsky01,pra04,science04,pra05}, this
however being just a sufficient and not necessary condition for
nonclassicality~\cite{lvovsky02:pra2}. It is still an open question which is the universal way to
experimentally characterize the nonclassicality of a quantum state.

A conceptually simple way to generate a quantum light state with a varying degree of nonclassicality
consists in adding a single photon to any completely classical one. This is quite different from photon
subtraction which, on the other hand, produces a nonclassical state only when starting from an already
nonclassical one~\cite{wenger04,kim05}.

In this Letter we report the generation and the analysis of single-photon-added thermal states (SPATSs),
i.e., completely classical states excited by a single photon, first described by Agarwal and Tara in
1992 \cite{agarwal92}. We use the techniques of conditioned parametric amplification recently
demonstrated by our group~\cite{science04,pra05} to generate such states, and we employ ultrafast pulsed
homodyne detection and quantum tomography to investigate their character. The peculiar nonclassical
behavior of SPATSs has recently triggered an interesting debate \cite{vogel00,diosi00} and has been
described in several theoretical papers~\cite{agarwal92,lee95,jones97,kim05,diosi00}; their experimental
generation has already been proposed, although with more complex schemes~\cite{jones97, kim05,dakna98},
but never realized. Thanks to their adjustable degree of quantumness, these states are an ideal
benchmark to test the different experimental criteria of nonclassicality recently proposed, and to
investigate the possibility of multi-photon entanglement generation. The nonclassicality of SPATSs is
here analyzed by reconstructing their negative-valued Wigner functions, by using the quadrature-based
Richter-Vogel (RV) criterion, and finally comparing these with two other methods based on quantum
tomography. In particular, we show that the so-called \textit{entanglement potential}~\cite{asboth05} is
a sensitive measurement of nonclassicality, and that it provides quantitative data about the possible
use of the states for quantum information applications in terms of the entanglement that they would
generate once sent to a 50-50 beam-splitter.

\section{Experimental}
The main source of our apparatus is a mode-locked Ti:Sa laser which emits 1.5 ps pulses with a
repetition rate of 82 MHz. The pulse train is frequency-doubled to 393 nm by second harmonic
generation in a LBO crystal. The spatially-cleaned UV beam then serves as a pump for a type-I BBO
crystal which generates spontaneous parametric down-conversion (SPDC) at the same wavelength of the
laser source. Pairs of SPDC photons are emitted in two distinct spatial channels called signal and
idler. Along the idler channel the photons are strongly filtered in the spectral and spatial domain
by means of etalon cavities and by a single-mode fiber which is directly connected to a
single-photon-counting module (further details are given in~\cite{pra04,pra05}). The signal field
is mixed with a strong local oscillator (LO, an attenuated portion of the main laser source) by
means of a 50\% beam-splitter (BS). The BS outputs are detected by two photodiodes connected to a
wide-bandwidth amplifier which provides the difference (homodyne) signal between the two
photocurrents on a pulse-to-pulse basis~\cite{josab02}. Whenever a single photon is detected in the
idler channel, an homodyne measurement is performed on the correlated spatio-temporal mode of the
signal channel by storing the corresponding electrical signal (proportional to the quadrature
operator value) on a digital scope.
\begin{figure}[h]
\includegraphics*[width=80mm]{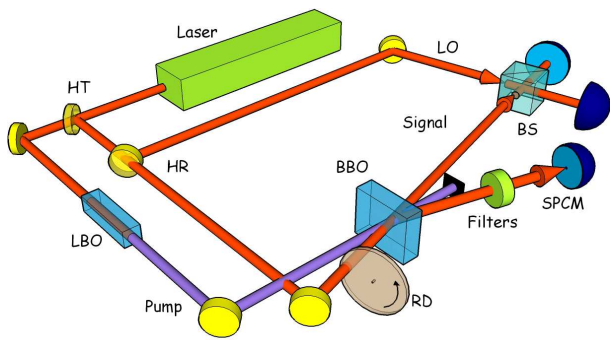}
\caption{(color online) Experimental setup. HR (HT) is a high reflectivity (transmittivity) beam
splitter; SPCM is a single-photon-counting module; all other symbols are defined in the text. The
mode-cleaning fiber used to inject the thermal state coming from the rotating ground glass disk
(RD) into the parametric crystal is not shown here for clarity. \label{fig:setup}}
\end{figure}

When no field is injected in the SPDC crystal, conditioned single-photon Fock states are generated
from spontaneous emission in the signal channel~\cite{lvovsky01,pra04}. We have recently shown
that, if the SPDC crystal is injected with a coherent state, stimulated emission comes into play
and single-photon excitation of such a pure state is obtained~\cite{science04,pra05}. However, a
coherent state is still at the border between the quantum and the classical regimes; it is
therefore extremely interesting to use a truly classical state, like the thermal one, as the input,
and to observe its degaussification~\cite{wenger04}. In order to avoid the technical problems
connected to the handling of a true high-temperature thermal source, we use pseudo-thermal one,
obtained by inserting a rotating ground glass disk (RD) in a portion of the laser beam (see
Fig.\ref{fig:setup}). By coupling a fraction (much smaller than the typical speckle size) of the
randomly scattered light into a single-mode fiber, at the output we obtain a clean spatial mode
with random amplitude and phase yielding the photon distribution typical of a thermal
source~\cite{arecchi65} which is then used to inject the parametric amplifier.

\section{Properties of SPATSs}
In order to describe the state generated in our experiment, we give a general treatment of photon
addition based on conditioned parametric amplification. By first-order perturbation theory, the
output of the parametric amplifier when a pure state $\ket{\varphi_m}$ is injected along the signal
channel is given by
\begin{equation}
\ket{\psi_m} = [1+(g\hat a^\dag_s\hat a^\dag_i - g^* \hat a_s\hat a_i)] \ket{\varphi_m}_s\ket{0}_i,
\end{equation}
where $g$ accounts for the coupling and the amplitude of the pump and $\hat a$, $\hat a^\dag$ are
the usual noncommuting annihilation and creation operators. For a generic signal input, the output
state of the parametric amplifier can be written as
\begin{equation}
\hat \rho_{\rm{out}} = \sum_m P_m \ket{\psi_m}\bra{\psi_m}
\end{equation}
where the input mixed state is $\hat \rho_s = \sum P_m \ket{\varphi_m}\bra{\varphi_m}$ and $P_m$ is the
probability for the state $\ket{\varphi_m}$. If we condition the preparation of the signal state to
single-photon detection on the idler channel, we obtain the prepared state
\begin{equation}
\hat \rho = {\rm{Tr}}_i(\hat \rho_{\rm{out}} \ket{1}_i\bra{1}_i) = |g|^2 \hat a^\dag_s \hat \rho_s \hat
a_s.
\end{equation}
When the input state $\hat \rho_s$ is a thermal state with mean photon number $\bar n$, we obtain
that the single-photon-added thermal state is described by the following density operator expressed
in the Fock base:
\begin{equation} \hat \rho = \frac{1}{\bar n (\bar n +1)}\sum_{n=0}^{\infty}{\left(
\frac{\bar n}{1+\bar n} \right)^n n \ket{n} \bra{n}}. \label{eq:spats:rho}
\end{equation}
The lack of the vacuum term and the rescaling of higher excited terms is evident in this expression. The
$P$ phase-space representation can be easily calculated and is given by (see also~\cite{agarwal92})
\begin{equation}
P(\alpha) = \frac{1}{\pi \bar n^3} [(1+\bar n)|\alpha|^2-\bar n] e^{-|\alpha|^2/\bar n},
\end{equation}
while the corresponding Wigner function reads as
\begin{equation}
W(\alpha)=\frac{2}{\pi} \frac{|2\alpha|^2(1+\bar n)-(1+2 \bar n)}{(1+2\bar n)^3} \
e^{-2|\alpha|^2/(1+2\bar n)} \label{eq:spats:wigner}
\end{equation}
where $\alpha = x + i y$. SPATSs have a well-behaved $P$ function which is always negative around
$\alpha = 0$; this feature is also present in the Wigner function and assures their nonclassicality,
however both $P(0)$ and $W(0)$ tend to zero in the limit of $\bar n \rightarrow \infty$.

\section{Data analysis and discussion}
\begin{figure}[h]
\includegraphics[width=85mm]{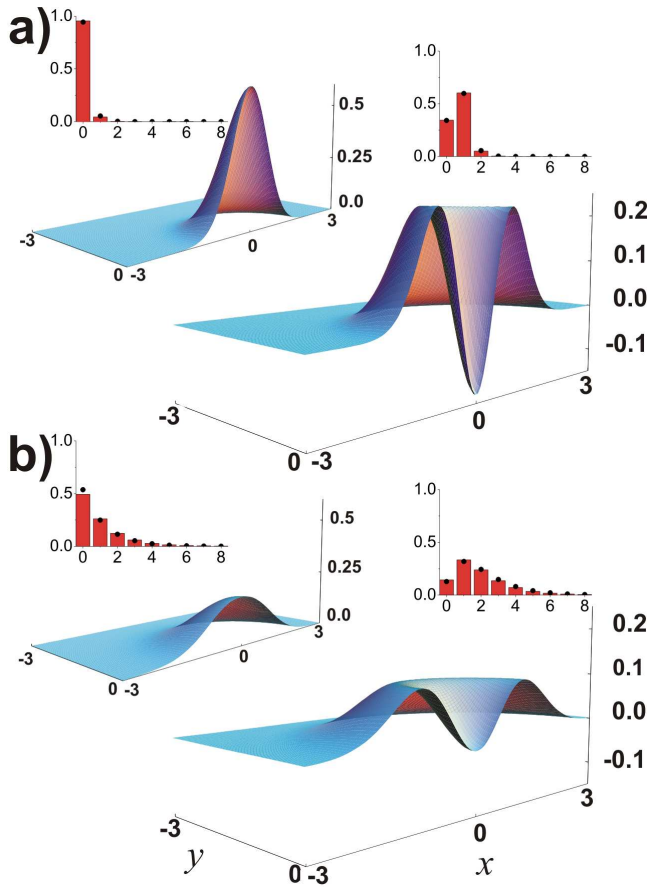}
\caption{(color online) Experimentally reconstructed diagonal density matrix elements (reconstruction
errors of statistical origin are of the order of $1\%$) and Wigner functions for thermal states (left)
and SPATSs (right): a) $\bar n = 0.08$; b) $\bar n = 1.15$. Filled circles indicate the density matrix
elements calculated for thermal states and SPATSs with the expected efficiencies.\label{fig:dm}}
\end{figure}

After the acquisition of about $10^5$ quadrature values with random phases, we have performed the
reconstruction of the diagonal density matrix elements using the maximum likelihood
estimation~\cite{banaszek99}. This method gives the density matrix that most likely represents the
measured homodyne data. Firstly, we build the likelihood function contracted for a density matrix
truncated to $25$ diagonal elements (with the constraints of Hermiticity, positivity and normalization),
then the function is maximized by an iterative procedure~\cite{lvovsky04,hradil06} and the errors on the
reconstructed density matrix elements are evaluated using the Fisher information~\cite{hradil06}. The
results are shown in Fig.~\ref{fig:dm}, together with the corresponding reconstructed~\cite{pra05}
Wigner functions
for two different temperatures of the injected thermal state. Since in the low-gain regime the
count rate in the idler channel is given by $\langle\hat n \rangle = \mathrm{Tr}(\hat
\rho_{\mathrm{out}} \hat a^\dag_i \hat a_i )=|g|^2(1+\bar n)$, the mean photon number values $\bar
n$ reported in Fig.~\ref{fig:dm} and in the following are obtained from the ratio between the
trigger count rates when the thermal injection is present and when it is blocked (see
Ref.~\cite{pra05} and references therein).

The finite experimental efficiency in the preparation and homodyne detection of SPATSs is fully
accounted for by a loss mechanism which can be modeled by the transmission of the ideal state $\hat
\rho$ of Eq.(\ref{eq:spats:rho}) through a beam splitter of trasmittivity $\eta$ coupling vacuum into
the detection mode, such that the detected state $\hat \rho_{\eta}$ is finally found as:
\begin{equation}
\hat \rho_{\eta}=\mathrm{Tr}_R \{U_{\mathrm{\eta}}(\hat\rho \ket{0}\bra{0})U^\dag_{\mathrm{\eta}}\}
\label{eq:spats:rhom}
\end{equation}
where $U_{\mathrm{\eta}}$ is the beam splitter operator acting on two input modes containing the state
$\hat \rho$ and the vacuum, and the states of the reflected mode (indicated by $R$) are traced out. In
the case of finite efficiency the expression for the Wigner function thus results:
\begin{equation}
W_{\eta}(\alpha)=\frac{2}{\pi} \frac{1+2\eta[\bar n +2(1+\bar n)|\alpha|^2-2\bar n \eta-1]}{(1+2\bar
n\eta)^3} \ e^{\frac{-2|\alpha|^2}{1+2\bar n\eta}}. \label{eq:spats:wigm}
\end{equation}

It should be noted that the value of experimental efficiency which best fits the data is the same ($\eta
= 0.62$) as that obtained for single-photon Fock states (i.e., without injection), and implies that only
a portion of vacuum due to losses enters the mode during the generation of SPATS. Thanks to a very low
rate of dark counts in the trigger detector, the portion of the injected thermal state which survives
the conditional preparation procedure and contributes to degradation of the SPATSs is in fact completely
negligible. However, since the nonclassical features of the state get weaker for large $\bar n$, a
limited efficiency ($\eta<1$) has the effect of progressively hiding them among unwanted vacuum
components.

\begin{figure}[h]
\includegraphics[width=80mm]{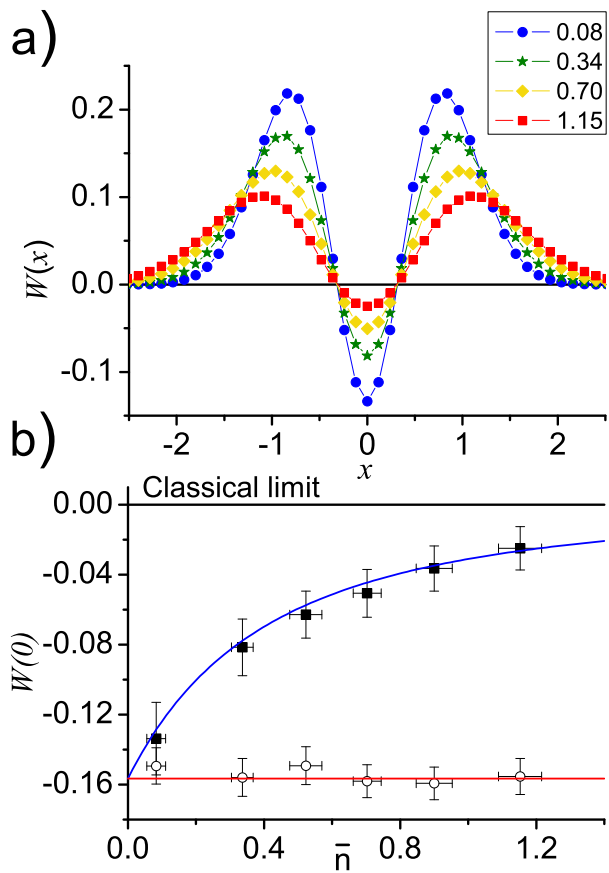}
\caption{(color online) a) Sections of the experimentally reconstructed Wigner functions for SPATSs with
different $\bar n$; b) Experimental values for the minimum of the Wigner function $W(0)$ as a function
of $\bar n$ for SPATSs (solid squares) and for single-photon Fock states (empty circles) obtained by
blocking the injection; the values calculated from Eq.(\ref{eq:spats:wigm}) for $\eta=0.62$ (solid
curves) are in very good agreement with experimental data and clearly show the appropriateness of the
model. Negativity of the Wigner function is a sufficient condition for affirming the nonclassical
character of the state.\label{fig:secwig}}
\end{figure}

\begin{figure*}[ht]
\includegraphics[width=140mm]{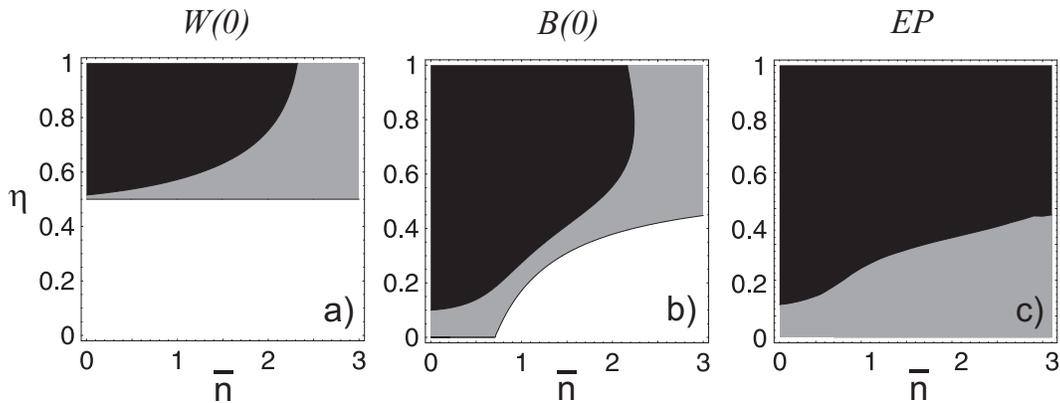}
\caption{Calculated regions of nonclassical behavior of SPATSs as a function of $\bar n$ and $\eta$
according to: a) the negativity of the Wigner function at the origin $W(0)$; b) the Klyshko criterion
$B(0)$; c) the entanglement potential EP. White areas indicate classical behavior; grey areas indicate
where a potentially nonclassical character is not measurable due to experimental reconstruction noise
(estimated as the average error on the experimentally reconstructed parameters); black areas indicate
regions where the nonclassical character is measurable given the current statistical
uncertainties.\label{fig:contour}}
\end{figure*}

Indeed, the measured negativity of the Wigner function at the origin (see Fig.\ref{fig:secwig}a and b)
rapidly gets smaller as the mean photon number of the input thermal state is increased. With the current
level of efficiency and reconstruction accuracy we are able to prove the nonclassicality of all the
generated states (up to $\bar n=1.15$), but one may expect to experimentally detect negativity above the
reconstruction noise, and thus prove state nonclassicality, up to about $\bar n \approx 1.5$ (also see
Fig.\ref{fig:contour}a). It should be noted that, even for a single-photon Fock state, the Wigner
function loses its negativity for efficiencies lower than $50\%$, so that surpassing this experimental
threshold is an essential requisite in order to use this nonclassicality criterion.

After having experimentally proved the nonclassicality of the states for all the investigated
values of $\bar n$, it is interesting to verify the nonclassical character of the measured SPATSs
also using different criteria.

The first one has been recently proposed by Richter and Vogel~\cite{vogel00} and is based on the
characteristic function $G(k,\theta)=\langle e^{ik\hat x(\theta)}\rangle$ of the quadratures (i.e.,
the Fourier transform of the quadrature distribution), where $\hat x(\theta)= (\hat a e^{-i\theta}
+ \hat a^{\dag}e^{i\theta})/2$ is the phase-dependent quadrature operator. At the first-order, the
criterion defines a phase-independent state as nonclassical if there is a value of $k$ such that
$|G(k,\theta)|\equiv|G(k)|
> G_{\rm gr}(k)$, where $G_{\rm gr}(k)$ is the characteristic function for the vacuum measured when the
signal beam is blocked before homodyne detection. In other words, the evidence of structures
narrower than those associated to vacuum in the quadrature distribution is a sufficient condition
to define a nonclassical state~\cite{lvovsky02:pra2}. However, it has been shown that nonclassical
states exist (as pointed out by Di\'osi~\cite{diosi00} for a vacuum-lacking thermal
state~\cite{lee95}, which is very similar to SPATSs) which fail to fulfil such inequality; when
this happens, the first-order Richter-Vogel (RV) criterion has to be extended to higher orders: the
second-order RV inequality reads as
\begin{equation}
2 G^2(k/2)G_{\rm gr}(k/\sqrt{2})-G(k) > G_{\rm gr}(k). \label{eq:second}
\end{equation}
It is evident that, as higher orders are investigated, the increasing sensitivity to experimental
and statistical noise may soon become unmanageable.

\begin{figure}[h]
\includegraphics[width=80mm]{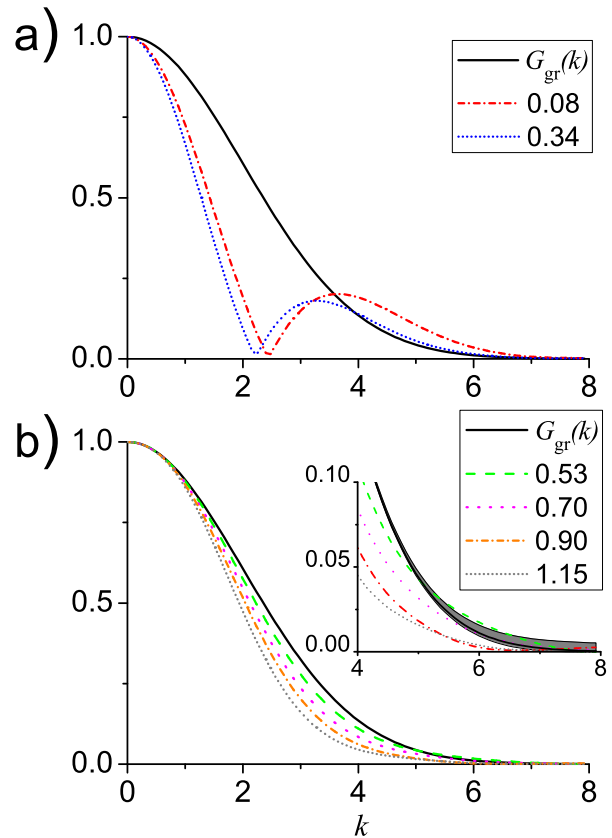}
\caption{(color online) Experimental characteristic functions involved in the RV nonclassicality
criterion for the detected SPATSs: a) first order; b) second order (the inset shows a magnified view of
the region where the state with $\bar n=0.53$ is just slightly fulfilling the criterion).
\label{fig:vogel}}
\end{figure}

The measured $|G(k)|$ and left hand side of Eq.~(\ref{eq:second}) are plotted in Fig.~\ref{fig:vogel}a)
and b), together with the $G_{\rm gr}(k)$ characteristic function, also obtained from the experimental
quadrature distribution of vacuum.
While the detected SPATSs satisfy the nonclassical first-order RV criterion only for the two lowest
values of $\bar n$, it is necessary to extend the criterion to the second order to just barely show
nonclassicality at large values of $k$ for $\bar n=0.53$ (see the inset of Fig.\ref{fig:vogel}b, where
the shaded region indicates the error area of the experimental $G_{\rm gr}(k)$).

At higher temperatures, no sign of nonclassical behavior is experimentally evident with this approach,
although the Wigner function of the corresponding states still clearly exhibits a measurable negativity
(see Fig.\ref{fig:secwig}). It should be noted that the second-order RV criterion for the ideal state of
Eq.~(\ref{eq:spats:rho}) is expected to prove the nonclassicality of SPATSs up to $\bar n \approx 0.6$
\cite{vogel00}; however, when the limited experimental efficiency and the statistical noise is taken
into account, it will start to fail even earlier.

\begin{figure}[h]
\includegraphics[width=80mm]{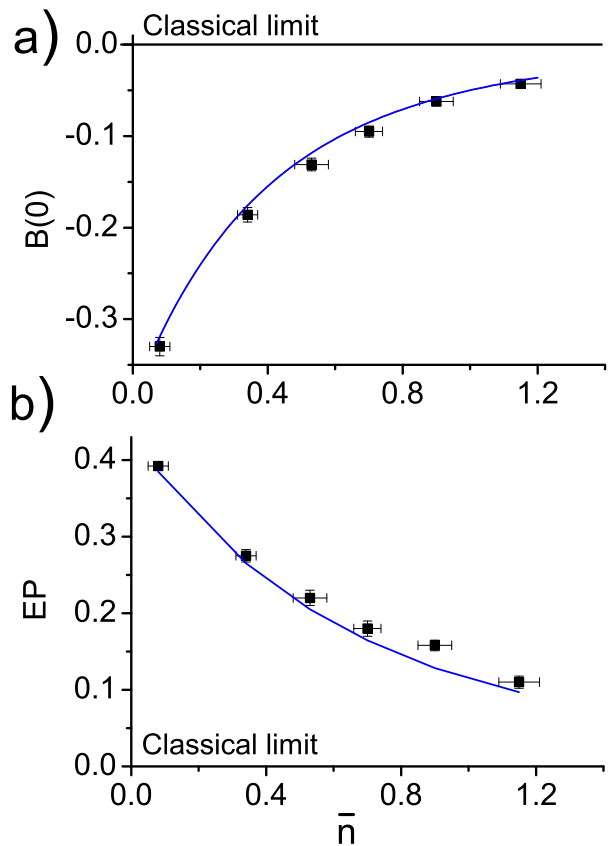}
\caption{(color online) a) Experimental data (squares) and calculated values (solid curve) of $B(0)$ as
a function of $\bar n$; negative values indicate nonclassicality of the state. b) The same as above for
the entanglement potential (EP) of the SPATSs; here nonclassicality is demonstrated by EP values greater
than zero. \label{fig:bep}}
\end{figure}

The tomographic reconstruction of the state that was earlier used for the nonclassicality test based on
the negativity of the Wigner function, can also be exploited to test alternative criteria: for example
by reconstructing the photon-number distribution $\rho_n = \bra{n}\hat \rho_{meas} \ket{n}$ and then
looking for strong modulations in neighboring photon probabilities by the following
relationship~\cite{klyshko96,dariano99}
\begin{equation}
B(n) \equiv (n+2) \rho_{n} \rho_{n+2}-(n+1)\rho_{n+1}^2<0,
\end{equation}
introduced by Klyshko in 1996, which is known to hold for nonclassical states. In the ideal situation of
unit efficiency SPATSs should always give $B(0)<0$ due to the absence of the vacuum term $\rho_{0}$, in
agreement with Ref.~\cite{lee95}. The experimental results obtained for $B(0)$ by using the
reconstructed density matrix $\hat \rho_{meas}$ are presented in Fig.\ref{fig:bep}a) together with those
calculated for the state described by $\hat \rho_{\eta}$ (see Eq.(\ref{eq:spats:rhom})) with
$\eta=0.62$. The agreement between the experimental data and the expected ones is again very
satisfactory, showing that our model state $\hat \rho_{\eta}$ well represents the experimental one. Our
current efficiency should in principle allow us to find negative values of $B(0)$ even for much larger
values of $\bar n$; however, if one takes the current reconstruction errors due to statistical noise
into account, the maximum $\bar n$ for which the corresponding SPATS can be safely declared nonclassical
is of the order of $2$. It should be noted that, differently from the Wigner function approach, here the
nonclassicality can be proved even for experimental efficiencies much lower than $50\%$, as far as the
mean photon number of the thermal state is not too high (see Fig.\ref{fig:contour}b).

Finally, it is particularly interesting to measure the entanglement potential (EP) of our states as
recently proposed by Asboth \textit{et al.}~\cite{asboth05}. This measurement is based on the fact that,
when a nonclassical state is mixed with vacuum on a 50-50 beam splitter, some amount of entanglement
(depending on the nonclassicality of the input state) appears between the BS outputs. No entanglement
can be produced by a classical initial state. For a given single-mode density operator $\hat\rho$, one
calculates the entanglement of the bipartite state at the BS outputs $\hat
\rho'=U_{\mathrm{BS}}(\hat\rho |0\rangle\langle 0|)U^\dag_{\mathrm{BS}}$ by means of the logarithmic
negativity $E_\mathcal{N}(\hat \rho')$ based on the Peres separability criterion and defined
in~\cite{vidal02}, where $U_{\mathrm{BS}}$ is the 50-50 BS transformation. The computed entanglement
potentials for the reconstructed SPATS density matrices $\hat \rho_{meas}$ are shown in
Fig.~\ref{fig:bep}b) together with those expected at the experimentally-evaluated efficiency (i.e.,
obtained from $\hat \rho_{\eta}$ with $\eta=0.62$). The EP is definitely greater than zero (by more than
$13 \sigma$) for all the detected states, thus confirming that they are indeed nonclassical, in
agreement with the findings obtained by the measurement of $B(0)$ and $W(0)$. As a comparison, the EP
would be equal to unity for a pure single-photon Fock state, while it would reduce to 0.43 for a
single-photon state mixed with vacuum $\hat \rho = (1-\eta)\ket{0}\bra{0}+\eta \ket{1}\bra{1}$ with
$\eta = 0.62$.

To summarize, the three tomographic approaches to test nonclassicality have all been able to
experimentally prove it for all the generated states (i.e., SPATSs with an average number of photons in
the seed thermal state up to $\bar n=1.15$) for a global experimental efficiency of $\eta=0.62$. In
order to gain a better view of the range of values for $\bar n$ and for the global experimental
efficiency $\eta$ which allow to prove the nonclassical character of single-photon-added thermal states
under realistic experimental conditions, we have calculated the indicators $W(0)$, $B(0)$, and EP from
the model state described by $\hat \rho_{\eta}$. The results are shown in Fig.\ref{fig:contour}: the
contour plots define the regions of parameters where the detected state is classical (white areas),
where it would result nonclassical if the reconstruction errors coming from statistical noise could be
neglected (grey areas) and, finally, where it is definitely nonclassical even with the current level of
noise (black areas).
From such plots it is evident that, as already noted, the Wigner function negativity only works for
sufficiently high efficiencies, while both $B(0)$ and EP are able to detect nonclassical behavior even
for $\eta < 50\%$. In particular, the entanglement potential is clearly seen to be the most powerful
criterion, at least for these particular states, and to allow for an experimental proof of
nonclassicality for all combinations of $\bar n$ and $\eta$, as long as reconstruction errors can be
neglected. Also considering the current experimental parameters, EP should show the quantum character of
SPATSs even for $\bar n>3$, thus demonstrating its higher immunity to noise.

Although at a different degree, all three indicators are however very sensitive to the presence of
reconstruction noise of statistical origin which may completely mask the nonclassical character of the
states, even for relatively low values of $\bar n$ or for low efficiencies. In order to unambiguously
prove the quantum character of higher-temperature SPATSs in these circumstances the only possibility is
to reduce the ``grey zone'' by significantly increasing the number of quadrature measurements.

\section{Conclusions}
In conclusion, we have generated a completely incoherent light state possessing an adjustable degree of
quantumness which has been used to experimentally test and compare different criteria of
nonclassicality. Although the direct analysis of quadrature distributions, done following the criterion
proposed by Richter and Vogel, has been able to show the nonclassical character of some of the states
with lower mean photon numbers, quantum tomography, with the reconstruction of the density matrix and
the Wigner function from the homodyne data, has allowed us to unambiguously show the nonclassical
character of all the generated states: three different criteria, the negativity of the Wigner function,
the Klyshko criterion and the entanglement potential, have been used with varying degree of
effectiveness in revealing nonclassicality. Besides being a useful tool for the measurement of
nonclassicality through the definition of the entanglement potential, the combination of nonclassical
field states - such as those generated here - with a beam-splitter, can be viewed as a simple entangling
device generating multi-photon states with varying degree of purity and entanglement and allowing the
future investigation of continuous-variable mixed entangled states~\cite{mhorodecki98}.

\section{Acknowledgments}
The authors gratefully acknowledge Koji Usami for giving the initial stimulus for this work and Milena
D'Angelo and Girish Agarwal for useful discussions and comments. This work was partially supported by
Ente Cassa di Risparmio di Firenze and MIUR, under the PRIN initiative and FIRB
contract RBNE01KZ94. 

\bibliography{Fock_bib}
\end{document}